\documentclass[fleqn,usenatbib]{mnras}

\usepackage{newtxtext,newtxmath}

\usepackage[T1]{fontenc}
\usepackage{ae,aecompl}

\usepackage{graphicx}	
\usepackage{amsmath}	
\usepackage{diagbox}
\usepackage{colortbl}
\definecolor{gray}{rgb}{0.6,0.6,0.6}

\newcommand{\dd}{\mathrm d}
\renewcommand{\d}{\partial}

\renewcommand{\(}{\left(}
\renewcommand{\)}{\right)}

\usepackage{soul}	
\usepackage[dvipsnames]{xcolor}

\title[Nonlinear particle reacceleration by multiple shocks]{Nonlinear particle reacceleration by multiple shocks}

\author[T. Vieu et al.]{
T. Vieu,$^{1}$\thanks{E-mail: vieu@apc.in2p3.fr}
S. Gabici,$^{1}$
V. Tatischeff$^{2}$
\\
$^{1}$Universit\'e de Paris, CNRS, Astroparticule et Cosmologie, F-75013 Paris, France\\
$^{2}$Universit\'e Paris-Saclay, CNRS/IN2P3, IJCLab, 91405 Orsay, France\\
}

\date{Accepted XXX. Received YYY; in original form ZZZ}

\pubyear{2021}

\begin{document}
\label{firstpage}
\pagerange{\pageref{firstpage}--\pageref{lastpage}}
\maketitle

\begin{abstract}
When the pressure of particles accelerated at shock waves is no longer negligible compared to the kinetic pressure of the gas, the linear theory of diffusive shock acceleration breaks down. This is expected in particular when the shock sweeps up preexisting cosmic rays, or when multiple shocks reaccelerate successively the same particles. To describe these systems, one has to account for the nonlinear backreaction of the particles on the magnetohydrodynamic flow. Using an up-to-date semi-analytical model of particle reacceleration at nonlinear shocks, we show that the presence of prexisting energetic particles strongly affects the shock profile, in such a way that the reacceleration of non thermal particles or the acceleration of particles from the thermal bath becomes less efficient. We further describe the evolution of the distribution of particles after several shocks and study the properties of the asymptotic solution. We detail the case of identical shocks as well as more realistic scenarios, including the heating of the medium or superbubble environments. When the particles are efficiently confined in the acceleration region, it is generally found that the spectrum converges toward a concave solution after a few tens of shocks, with a spectral index around 3.5 at the highest energy. The postshock cosmic ray pressure reaches an asymptotic value of about 4~--~5\% of the ram pressure of one shock. Most of the shock pressure is transferred to escaping particles.
\end{abstract}




\section{Introduction}
The main sources of cosmic rays (CR) in our galaxy are believed to be supernova remnant (SNR) shocks. Particle acceleration at shocks has been investigated for decades under the framework of diffusive shock acceleration (DSA) \citep{axford1977,krymskii1977,bell1978,blandford1978}. A standard simplifying assumption made in DSA computations is that the accelerated CRs are not energetic enough to influence the fluid dynamics. This gives rise to the theory of linear DSA, which has the remarkable property of predicting a universal power-law in momentum for the distribution function of CR accelerated at strong shocks ($f(p) \propto p^{-4}$).

The linear theory is however expected to break down in most astrophysical environments. The CR luminosity of the galaxy indeed suggests that a fraction of about 10\% of the energy provided by supernovae (SNe) is converted into CRs \citep{strong2010}. This requires high injection efficiencies at the shock, such that the CR pressure cannot be neglected anymore.
Besides, CR are in equipartition with the gas pressure in the interstellar medium: not only the shock accelerates fresh CRs, it also sweeps up preexisting non thermal particles (hereafter called seeds). When SNRs expand in low density media, they span a large volume, such that the total energy of the seeds may become of the order of $10^{51}$ erg, that is, the energy of the SN explosion. In these cases, an accurate DSA computation requires to consider nonlinear effects.

Since the pioneer works of \citet{eichler1979,blandford1980,drury1981}, there have been numerous attempts to model nonlinear particle acceleration at shocks, including Monte-Carlo simulations \citep{ellison1984}, two-fluid models \citep[e.g.][]{malkov1996}, semi-analytical solutions \citep{blasi2002,caprioli2009}, numerical resolutions \citep[e.g.][]{kang2011}, hydrodynamic simulations \citep[e.g.][]{caprioli2014}. It is now understood that the backreaction of the accelerated particles creates a precursor in the gas upstream of the shock. The compression factor at the subshock decreases while the total compression factor increases, in such a way that the CR spectrum tends to develop a concave shape instead of the universal power law predicted by the linear theory.

All the aforementioned works assumed that no seeds preexist in the medium before the acceleration process, i.e., the shock is isolated. Multiple shock acceleration is however expected to take place in several physical environments, including, e.g., stellar winds \citep{white1985}, accretion disks \citep{spruit1988,achterberg1990}, superbubbles \citep{bykov2001,parizot2004}, or clusters of galaxies \citep{gabici2003,kang2021b}. Within the framework of linear DSA, it was already shown by \citet{bell1978b} that multiple shock acceleration efficiently produce hard spectra. Including adiabatic decompression, \citet{melrose1993} demonstrated that the spectrum tends toward the asymptotic solution $f(p) \propto p^{-3}$. Nonlinear effects were tackled in \citet{blasi2004}, where it was shown that the flow profile could be indeed strongly affected by the seeds. \citet{kang2011} also studied such effects in the context of weak cosmological shock waves. As for the reacceleration of particles by an ensemble of nonlinear shocks, it has been very scarcely analysed in previous works. Considering the closely related situation of CR acceleration by an ensemble of coexisting shocks, \citet{bykov2001} included the backreaction of the CRs onto the shocks within a stochastic approach. Although this is a self-consistent way to account for the nonlinearities and maintain the energy balance between the shocks and the particles, it does not account for the local changes in the shock profiles due to the CR pressure. The problem of the acceleration by successive nonlinear shocks was later briefly investigated numerically by \citet{ferrand2008}, and more recently by \citet{caprioli2018}, who showed that the seeds are in particular expected to enhance streaming instabilities.

In this paper, we aim at computing the spectrum of reaccelerated seeds as well as the acceleration by successive shocks of particles from the thermal bath, in the nonlinear regime. In Section~\ref{sec1}, we describe an up-to-date version of the semi-analytical model of \citet{blasi2004}, including more recent developments describing injection, streaming instability and Alfvénic drift. In Section~\ref{sec2}, we discuss the reacceleration of preexisting seeds. Particle acceleration by an ensemble of shocks is tackled in Section~\ref{sec3}. We conclude in Section~\ref{secconclusion}. 


\section{Nonlinear diffusive shock reacceleration}\label{sec1}
\subsection{Kinetic equation}
We consider an infinite and plane shock characterised by a Mach number $M_0$, propagating along the $x$ axis into a region filled by a population of non-thermal particles (seed particles) whose distribution function is spatially homogeneous far upstream and denoted by $f_\infty(p)$, where $p$ is the particle momentum. 
The shock also accelerates particles from the thermal gas. 
These freshly accelerated particles are injected in the accelerator at the location of the shock ($x=0$). 
In the following, the indices $i = $ 0,1,2 will refer to quantities (e.g. the fluid velocity $u_i$) at upstream infinity, immediately upstream of the shock, and immediately downstream of the shock, respectively. 
We further define the total shock compression factor $R_{\rm tot} = u_0/u_2$ and the compression factor at the subshock $R_{\rm sub} = u_1/u_2$. 
The stationary transport equation for the distribution function of accelerated particles $f(x,p)$ reads, in a reference frame where the shock is at rest \citep{drury1983}:
\begin{equation}\label{transportequation}
    \d_x \( D \d_x f \) - u \d_x f + \frac{p}{3} \frac{du}{dx} \d_p f + Q_1 \delta(x) \delta(p- p_0) = 0
    \, ,
\end{equation}
with the boundary condition $f(x=-\infty,p) = f_{\infty}(p)$. 
Here, $D$ is the particle diffusion coefficient and $Q_1$ the injection rate of particles from the thermal gas into the acceleration process, at the injection momentum $p_0$.

Energetic particles are confined at shocks due to repeated scattering off MHD (e.g. Alfven) waves, that keep the particle distribution function very close to isotropy.
Upstream of the shock a small anisotropy appears ($f(x,p)$ is not spatially uniform) and this triggers the CR streaming instability, resulting in the growth of Alfven waves \citep{bell1978}.
Only Alfven waves propagating in the direction of the stream of particles grow, i.e. those moving towards the negative $x$ direction. 
This implies that upstream of the shock the fluid velocity $u$ that appears in Eq.~\ref{transportequation} should be substituted by the velocity of the scattering centres, $u - v_A$ where $v_A$ is the Alfven speed.
At equilibrium, the particle distribution function downstream of the shock is spatially uniform, and therefore streaming instability does not operate there.
Moreover, if the magnetic turbulence is isotropised after the passage through the shock, the effective Alfven speed vanishes and scattering centres move away from the shock at the fluid speed $u_2$.
In the presence of strong field amplification the substitution $u \rightarrow u-v_A$ may impact significantly onto the spectrum of accelerated particles, as first noticed in \citet{zirakashvili2008} and \citet{caprioli2012}\footnote{In fact, a non vanishing values of the Alfven speed might also be present downstream, due to the {\it inertia} of waves excited upstream and compressed by the shock \citep{caprioli2020}. Such an effect, not considered here, would further increase the impact that the drift of scattering centres has on the spectrum of CRs accelerated at the shock.}. 

The solution of Eq.~\ref{transportequation} can be found by integrating it first between $x = 0^-$ and $x = 0^+$ and then between $x = -\infty$ and $x = 0^-$. 
The following differential equation is obtained:
\begin{equation}\label{transportequationatshock}
    \frac{p}{3} \(u_2 - u_p \) \frac{\dd f_1}{\dd p} = \( u_p + \frac{p}{3} \frac{\dd u_p}{\dd p} \) f_1 - u_0 f_\infty - Q_1 \delta (x) \delta(p-p_0)
    \, ,
\end{equation}
where we introduced the quantity $u_p$ defined as \citep{blasi2002}:
\begin{align}\label{Up}
    &u_p(p) \equiv u_1-v_{A,1} - \frac{1}{f_1(p)}\int_{-\infty}^0 \dd x \, \d_x \(u - v_A \) f(x,p)
    \, .
\end{align}
When seed particles are neglected, $u_p$ represents the characteristic velocity of scattering centres experienced upstream of the shock by particles of momentum $p$. When seeds are taken into account, the physical meaning of $u_p$ is not as straightforward. 
Yet, it remains a useful mathematical quantity to carry out further computations. 

Following here the phenomenological approach presented in \citet{blasi2005}, we constrain the injection term in Eq.~\ref{transportequation} in such a way to guarantee the continuity between the momentum distribution of thermal particles downstream of the shock (the Maxwell-Boltzmann distribution) and that of accelerated ones. This approach, admittedly oversimplified, is broadly consistent with results coming from sophisticated numerical simulations of shocks \citep{caprioli2014}. Moreover, in the following we will assume that seed particles with momentum smaller than $p_0$ are thermalised once they cross the shock. This allows to set the following boundary condition in momentum:
\begin{equation}\label{f1p0}
f_1(p_0) = \frac{n_0 R_{\rm tot}}{\pi^{3/2} p_0^3} \xi^3 e^{-\xi^2} \, ,
\end{equation}
where $n_0$ is the gas density far upstream of the shock and $\xi$ is the injection parameter such that $p_0 = \xi p_{th,2} = \xi \sqrt{2 m_p k T_2}$, with $m_p$ the proton mass (assuming a hydrogen gas) and $T_2$ the downstream temperature. This parametrisation is useful to give a more physical meaning to the injection momentum, as discussed in \citet{blasi2005}.

At this point, before describing the procedure used to solve Eq.~\ref{transportequationatshock}, it is mandatory to discuss how the shock transition is modified in the presence of a non-negligible pressure carried by CR particles.

\subsection{Fluid equations}

The fluid dynamics of the shock transition is governed by the mass and momentum conservation laws:
\begin{align}
    &\rho_0 u_0 = \rho(x) u(x) \, ,
\\
& \rho_0 u_0^2 + p_{g,0} + p_{c,0} = \rho(x) u(x)^2 + p_{g}(x) + p_{c}(x) + p_{B}(x) \, ,
\end{align}
where $\rho$ and $u$ represent the gas density and velocity at a given position $x$ upstream of the shock, and $p_g$, $p_c$, and $p_B$ the pressure of gas, CR, and magnetic field, respectively. We assume that the background magnetic field can be neglected compared to the amplified field.
It is convenient to divide the momentum equation by $\rho_0 u_0^2$ and introduce normalised pressures ($P_g = p_g/\rho_0 u_0^2$, etc.) to obtain:
\begin{equation}\label{momentumconservation}
    1 + P_{g,0} + P_{c,0} = U(x) + P_{g}(x) + P_{c}(x) + P_{B}(x) \, ,
\end{equation}
where $U(x) \equiv u(x)/u_0$. 
Assuming an adiabatic equation of state for the gas in the upstream region with adiabatic index $\gamma$ we can write:
\begin{equation}
    P_g(x) = \frac{U^{-\gamma}(x)}{\gamma M_0^2} ~ .
\end{equation}
The CR pressure is related to the particle distribution function $f(x,p)$ as:
\begin{equation}
    P_c(x) = \frac{4 \pi}{3 \rho_0 u_0^2} \int_{p_0}^\infty \dd p p^3 v(p) f(x,p) ~ .
\end{equation}
Finally, the spatial variation of the magnetic pressure ahead of the shock has been computed by \citet{caprioli2012}, under the assumption that the magnetic field is amplified due to CR streaming upstream of the shock.
This gives the expression:
\begin{equation}\label{magneticpressure}
    P_B(x) = \frac{2}{25} \frac{\( 1 - U(x)^{5/4} \)^2}{U(x)^{3/2}}
    \, ,
\end{equation}
which is accurate at the second order in $v_A/u$. 
Formally, this equation has been derived for a shock characterized by very large Mach and Alfvenic numbers, $M_0^2 \gg 1$ and $M_A^2 \gg 1$.

\subsection{Method of solution}
The solution of the problem is obtained by solving the transport equation for CR (Eq.~\ref{transportequationatshock}), coupled with the momentum conservation equation for the shock transition (Eq.~\ref{momentumconservation}).
This can be done in an approximate but still accurate way by introducing a distance $x_p(p)$ upstream of the shock defined in this way: particles accelerated to a momentum $p$ can probe a region ahead of the shock up to a distance $x_p(p)$. This can be expressed mathematically as:
\begin{equation}
f(x,p) = \left[ f_1(p) - f_{\infty}(p) \right] \vartheta[x-x_p(p)] + f_{\infty}(p)
\, ,
\end{equation}
as first pointed out by \citet{eichler1979} and later extensively used in the literature \citep[e.g.][and references therein]{blasi2002,amato2008}.

After adopting this assumption, the expression for the CR pressure at a given position simplifies significantly and can be written as\footnote{This expression slightly differs from Eq.~15 in \citet{blasi2004}. Even though the approach used in \citet{blasi2004} is not fully consistent, their results are not expected to be strongly impacted.}:
\begin{multline}
    P_c(x_p) \approx \frac{4 \pi}{3 \rho_0 u_0^2} \left\{ 
    \int_{p_0}^p \dd p' p'^3 v(p') f_\infty(p')
    \right. \\ \left.
    + \int_p^{\infty} \dd p' p'^3 v(p') f_1(p')
    \right\}
    \, .
\end{multline}
Moreover, Eq.~\ref{Up} becomes:
\begin{align}\label{simplifyUp}
    U_p(p) \approx \( U(x_p) - V_A(x_p) \) \( 1 - \frac{f_\infty(p)}{f_1(p)} \) + \frac{f_\infty(p)}{f_1(p)}  \, .
\end{align}
Solving this equation gives $U(x_p)$ as function of $p$. For the sake of clarity, we make this dependency explicit by renaming $U(x_p)$ as $\zeta(p)$. Using Eq.~\ref{magneticpressure} in order to express the Alfven velocity as function of the fluid velocity, Eq.~ \ref{simplifyUp} leads to:
\begin{equation}
    U_p(p) \approx \(\frac{7}{5} \zeta(p) - \frac{2}{5} \zeta(p)^{-1/4} \) \( 1 - \frac{f_\infty(p)}{f_1(p)} \) + \frac{f_\infty(p)}{f_1(p)}
    \, .
\end{equation}
Plugging this expression into Eq.~\ref{transportequationatshock} provides, after some algebra:
\begin{multline}\label{transportequationatshockworkedout}
    \frac{p}{3} \frac{\dd f_1}{\dd p} \(
    \frac{1}{R_{\rm tot}} - \frac{7}{5} \zeta + \frac{2}{5} \zeta^{-1/4} \)
    =  \\
    \frac{f_1 - f_\infty}{5} \( 
     7 \zeta - 2 \zeta^{-1/4}
     + \frac{p}{6} \(14 + \zeta^{-5/4} \) \zeta'(p) \) 
     \, .
\end{multline}

\begin{figure*}
\centering
  \includegraphics[width=\linewidth]{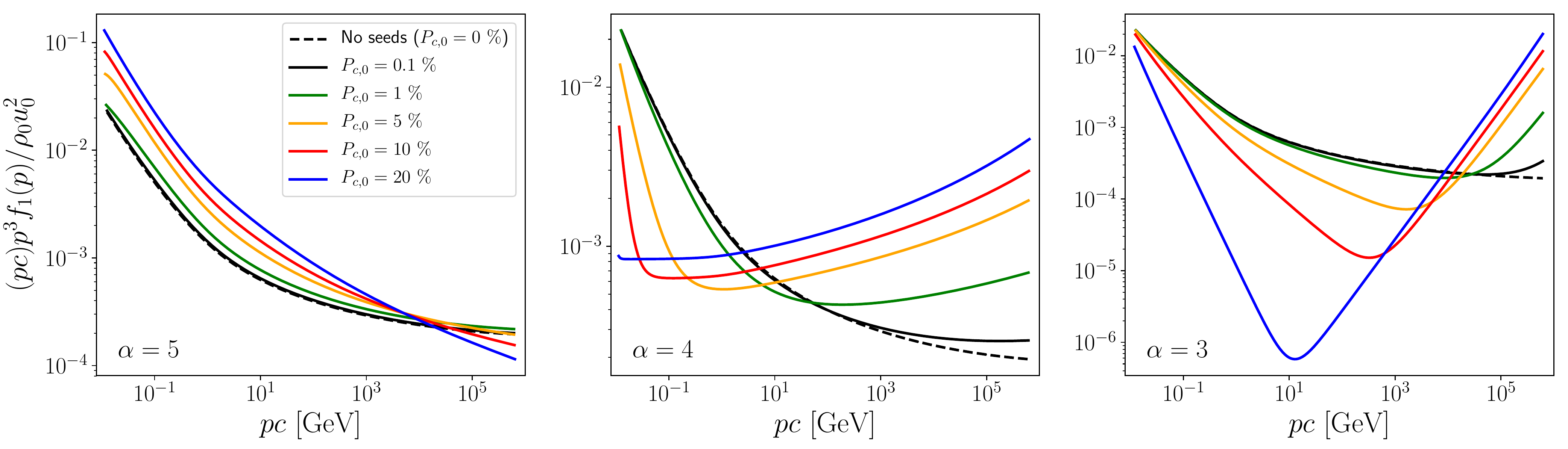}
\caption{Spectra after reacceleration and decompression of seeds with spectral index $\alpha=5$ (left), $\alpha=4$ (middle), $\alpha=3$ (right). Each curve corresponds to a different value of the relative CR pressure at upstream infinity.}
\label{fig:reacceleration_seeds}
\end{figure*}

The fluid and magnetic pressure terms evaluated at $x=x_p$ are functions of $U(x_p)$ only, while the CR pressure term at $x_p$ is function of $U(x_p)$ and $p$. Evaluating the momentum equation at $x_p$, we therefore get an equation which only depends on $\zeta(p)$ and $p$. After differentiating this equation with respect to $p$, we get:
\begin{multline}\label{momentumdifferential}
    \zeta'(p) \left[ \frac{27}{25} - \frac{\zeta^{-\gamma-1}}{M_0^2} + \frac{\zeta^{-5/4}}{25} - \frac{3 \zeta^{-5/2}}{25} \right]
    \\
    = \frac{4 \pi}{3 \rho_0 u_0^2} p^3 v(p) \left[
    f_1(p) - f_\infty(p)
    \right]
    \, .
\end{multline}
Two boundary conditions are needed to solve Eq.~\ref{transportequationatshockworkedout} together with  Eq.~\ref{momentumdifferential}. Eq.~\ref{f1p0} provides the boundary condition for the distribution function.
Then, we start with an initial guess value of $U_1$, which provides an initial value for $\zeta$, as $\zeta(p_0) \approx U_1$. The total compression factor is computed from the Rankine-Hugoniot condition at the subshock including the dynamical effect of the magnetic field \citep{caprioli2009}:
\begin{align}
\begin{aligned}
    &R_{\rm tot}^{\gamma+1} = \frac{M_0^2 R_{\rm sub}^\gamma}{2} \( \frac{\gamma+1 - R_{\rm sub} (\gamma-1)}{1+\Lambda_B} \)
    \, ,
    \\ \label{RHlamb}
    &\Lambda_B = \frac{2}{25} \(1-U_1^{5/4}\)^2 U_1^{\gamma-3/2} \gamma M_0^2 \( 1+ R_{\rm sub} \(\frac{2}{\gamma}-1 \) \)
    \, .
\end{aligned}
\end{align}

Now that two initial values have been obtained for the functions $f$ and $\zeta$, together with the properties of the subshock, Eqs.~\ref{momentumdifferential} and \ref{transportequationatshockworkedout} can be solved together numerically as follows. Eq.~\ref{momentumdifferential} gives $\zeta'(p)$ and $\zeta(p+dp)$ as function of $f_1(p)$ and $\zeta(p)$. Eq.~\ref{transportequationatshockworkedout} gives $f_1(p+dp)$ as function of $f_1(p)$, $\zeta(p)$ and $\zeta'(p)$. We can therefore reconstruct the full solutions $f_1$ and $\zeta$ for a given guess value of $U_1$. The physical value of $U_1$ is the one for which $\zeta(p_{\rm max}) = 1$, as the flow profile should not be modified at large distances from the shock. The determination of the maximum momentum $p_{\rm max}$ is not straightforward as it requires in particular to account for the time-dependency of the acceleration process. The maximum momentum $p_{\rm max}$ is left as a free parameter in this work, together with the Mach number of the shock $M_0$ and the injection parameter $\xi$.

\subsection{Adiabatic decompression}
The particles bound to the fluid are compressed by a factor $R_{\rm tot}$ after the passage of a shock. When successive reaccelerations occur, the decompression of the particles must be computed in between each shock, otherwise one could obtain arbitrarily hard spectra at low energy. An accurate computation of this decompression makes use of Liouville's theorem: $f_{\rm decompressed}(p) = f_{1}(R_{\rm tot}^{1/3} p)$ \citep{melrose1993}. As pointed out by \citet{ferrand2008}, in doing numerical resolutions the steps of the momentum grid in logarithmic scale should be chosen as exact fractions of the momentum shift $\log(R_{\rm tot})/3$. In nonlinear shocks, this requirement cannot be fulfilled since the total compression factor is unknown a priori. This introduces numerical errors in the decompressed spectrum. With the momentum resolution used to obtain our results,
the relative error between the expected pressure and that obtained numerically is equal to 5\% in the worst cases. The overall shape of the CR spectrum is therefore expected to stay accurate. However, when dealing with multiple successive shocks, it is crucial to correct the postshock pressure, otherwise the errors will accumulate. We therefore slightly rescale the decompressed spectrum in order to ensure that an adiabatic change takes place, imposing that $P_{c,{\rm decompressed}} = R_{\rm tot}^{-\gamma_c} P_{c,1}$, where $\gamma_c$ is the CR adiabatic index and $P_{c,1}$ the CR pressure at the shock. The renormalisation factor is equal to 0.95 in the worst cases.

In the following, the decompressed distribution of CRs remaining after the passage of a shock will be referred to as the ``postshock'' distribution.

\subsection{Escape flux}
Because modified shocks usually produce spectra harder than $p^{-4}$ at high energies, a non negligible amount of energy carried by escaping particles leaks upstream of the flow. The escape flux $F_{e}$ normalised to the kinetic energy of the shock can be computed using the conservation of the energy between the downstream region and upstream infinity \citep{blasi2005}:
\begin{align}\label{DSA:escapeflux}
	&F_e = 1 - \frac{1}{R_{\rm tot}^2} + \frac{2}{M_0^2 (\gamma-1)} - \frac{2}{R_{\rm tot}} \frac{\gamma}{\gamma-1} P_{g,2}
	\\ & \qquad \qquad \qquad  + \frac{2}{R_{\rm tot}} \frac{\gamma_c}{\gamma_c-1} \( P_{c,0} - \frac{P_{cr,2}}{R_{\rm tot}} \) \notag
	\, ,
	\\
	&P_{g,2} = U_1 - \frac{1}{R_{\rm tot}} + \frac{1}{\gamma M_0^2} U_1^{-\gamma} + \frac{2}{25 U_1^{3/2}} \( 1- U_1^{5/4} \)^2 \, ,
\end{align}
where $\gamma_c$ is the adiabatic index of the particles.
The term $P_{c,0}$ accounts for the pressure of the seeds at upstream infinity.

\section{Reacceleration of preexisting cosmic rays} \label{sec2}

In this section we investigate the effect of the nonlinearities on the reacceleration of preexisting seeds.
We set the shock Mach number $M_0=20$ ($u_0 = 3320$~km/s), the density far upstream $n_0=0.01$~cm$^{-3}$, the temperature $T_0 = 10^6$~K, the adiabatic index of the gas $\gamma=5/3$, the injection parameter $\xi=3$ and the maximum momentum $p_{\rm max}=1$~PeV. The spectrum of the seeds is assumed to be a power law of index $\alpha$.

Figure~\ref{fig:reacceleration_seeds} shows the spectra resulting from the acceleration of seeds with spectral indices 3 (hard), 4 (flat) and 5 (steep).
In the absence of seeds, the spectrum is moderately concave, with a change of slope around 10~GeV.
For seed spectra steeper than $p^{-4}$, the energy of the seeds is initially located around the injection momentum and there is not much difference between the reacceleration of these seeds and the acceleration of particles from the thermal bath, as can be seen in the left panel of Figure~\ref{fig:reacceleration_seeds}. In contrast, if the spectral index of the seeds is ``flat'' ($\alpha = 4$), the plasma flow is much more modified compared to the case where seeds are not present. As the compression factor of the subshock decreases, the injection of particles from the thermal pool becomes inefficient. On the other hand, high energy seeds feel the total compression factor and can thus be efficiently reaccelerated, which hardens the spectra at high energies. This results in hard spectra over almost all energy bands, as can be seen in the middle panel of Figure~\ref{fig:reacceleration_seeds}. Finally, if the spectrum of the seeds is harder than $p^{-4}$, the reacceleration becomes inefficient at high energies such that the high energy end of the spectrum is close to that of the seeds. On the other hand, the freshly injected particles are not efficiently accelerated at the subshock, which leads to extremely steep spectra at low energies. This results in very concave spectra, which are basically the superposition of an injection with the preexisting spectrum, as seen in the right panel of Figure~\ref{fig:reacceleration_seeds}.

\begin{figure}
\centering
  \includegraphics[width=\linewidth]{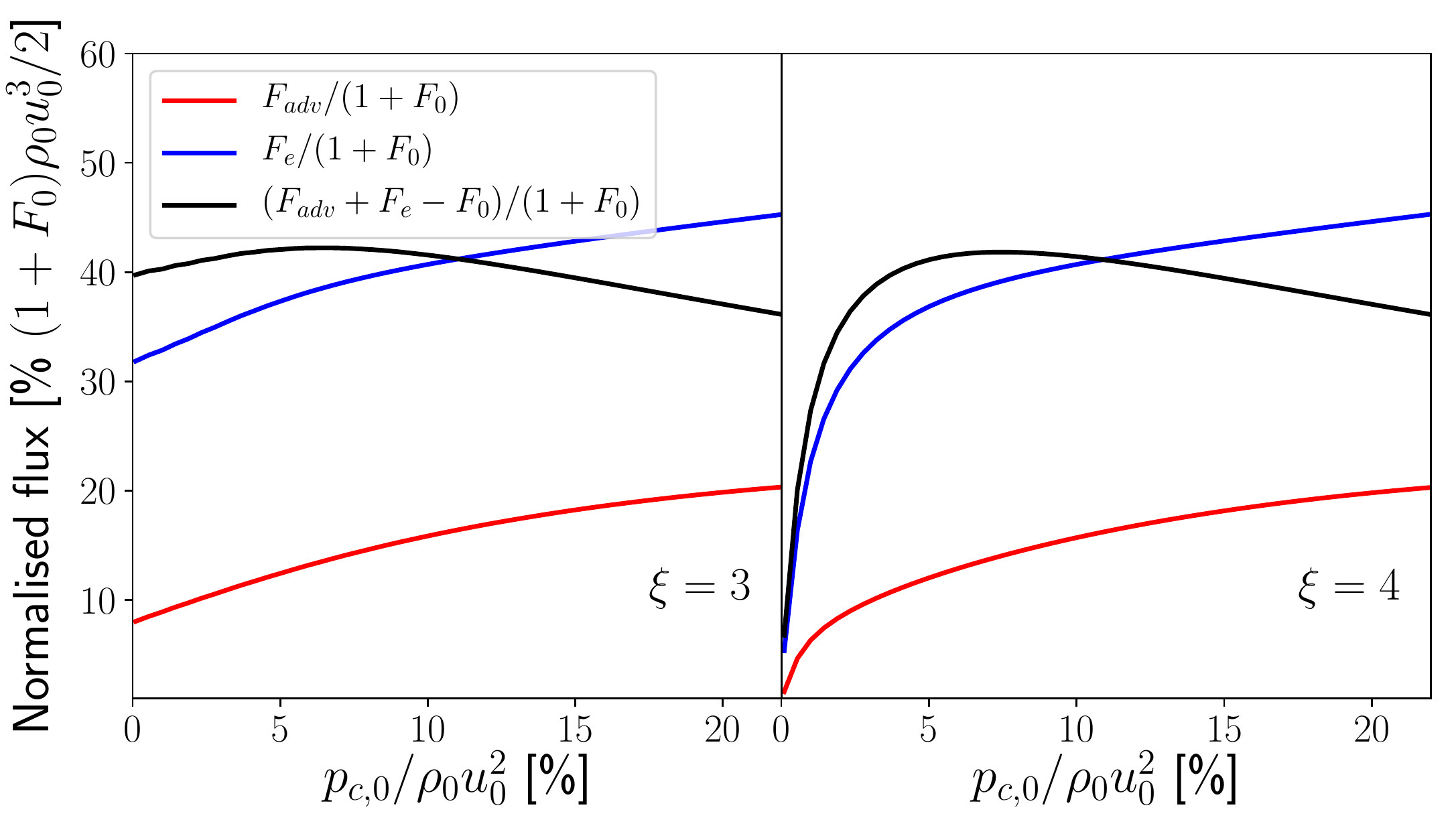}
\caption{Flux advected downstream $F_{\rm adv}$, escape flux upstream $F_{e}$, net flux gain $F_{\rm adv} + F_{e} - F_0$, all normalised to the upstream incoming flux (ram hydrodynamical flux $\rho u_0^3/2$ plus flux of preexisting particles $F_0$), as function of the pressure of the seeds at upstream infinity, for an injection parameter $\xi=3$ (left) and $\xi=4$ (right).}
\label{fig:reacceleration_seeds_flux}
\end{figure}

\begin{figure*}
\centering
  \includegraphics[width=0.8\linewidth]{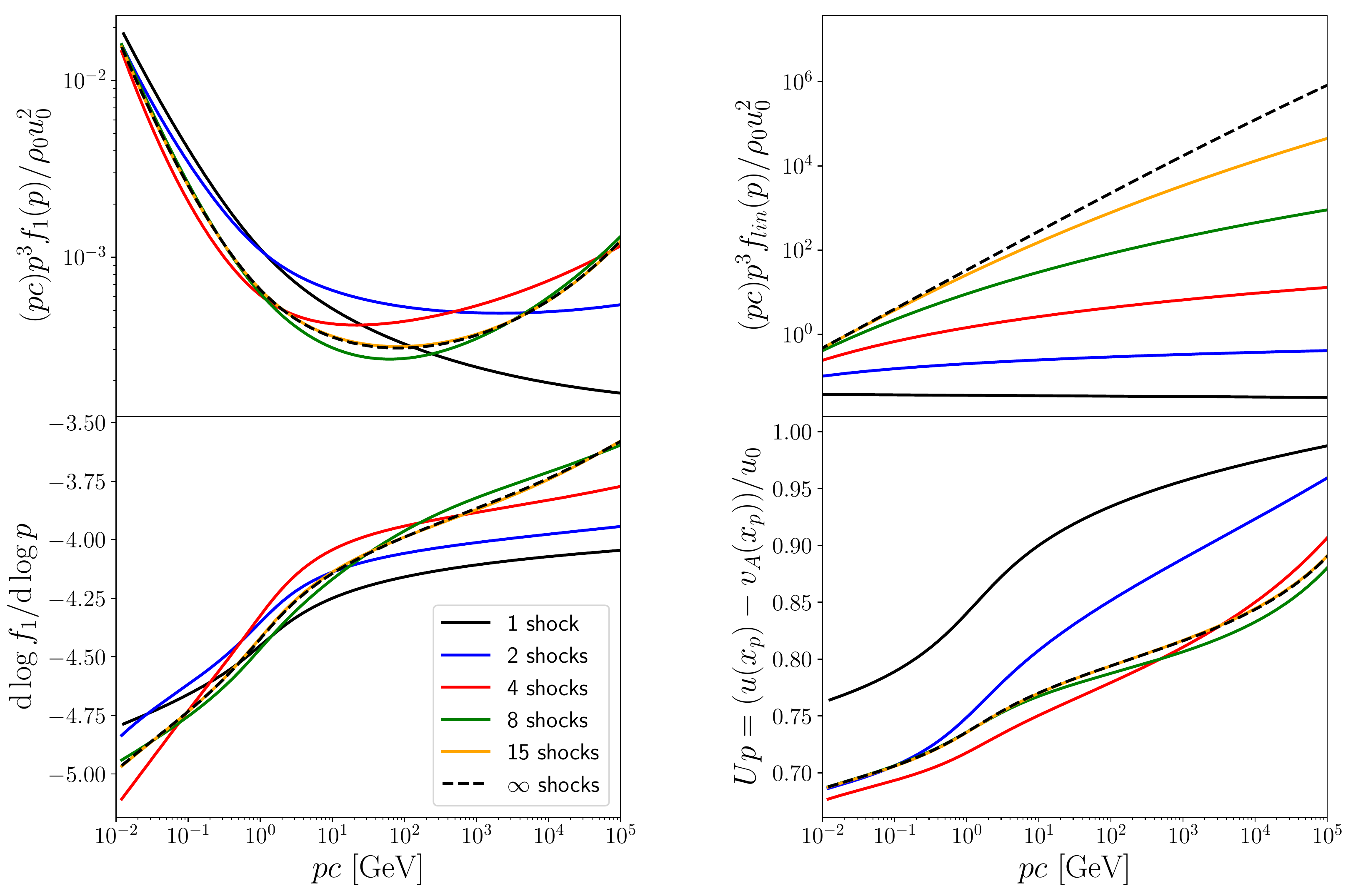}
\caption{CR spectrum after multiple reaccelerations (top left) and corresponding spectral indices (bottom left) compared with linear diffusive shock reacceleration (top right). The bottom right panel displays the velocity felt by the particles as defined by Eq.~\ref{Up}.}
\label{fig:successive_shocks}
\end{figure*}

Figure~\ref{fig:reacceleration_seeds_flux} shows how the CR flux is shared between the different regions, for a seed spectrum scaling as $p^{-4}$. Besides the escaping flux $F_e$, also shown are the flux advected downstream $F_{\rm adv}$ as well as the net flux gained in the acceleration $F_{\rm adv} + F_e - F_0$, where $F_{\rm adv}$ and $F_0$ are identified in Eq.~\ref{DSA:escapeflux} as:
\begin{align}\label{multipleshocks:fluxdownstream0}
F_{\rm adv} &= \frac{2}{R_{\rm tot}^2} \frac{\gamma_c}{\gamma_c-1} P_{cr,2} \, ,
\\
F_0 &= \frac{2}{R_{\rm tot}} \frac{\gamma_c}{\gamma_c-1} P_{c,0}
\, .
\end{align}
Although the energy flux would rapidly grow to unphysical values in the test-particle regime, accounting for the nonlinearity of the problem leads to a drastic reduction of the acceleration efficiency. The energy gain saturates at the level of about 40\% and then decreases as the pressure of the seeds is further increased, for in this case most of the flux of preexisting particles is converted into kinetic shock modification and heat. The upstream escaping flux is always a few times larger than the flux advected downstream, which is expected since the solution is a concave spectrum.


\section{Particle acceleration by successive nonlinear shocks}
\label{sec3}

\subsection{Identical shocks}\label{sec:multipleshocks_identical}
We now aim at investigating the acceleration of particles by multiple shocks. First we assume that all shocks are identical. This is a simplistic modelling as we expect for instance the medium to be heated by each shock if they all span the same volume, or the density to decrease if the volume expands. The idealistic solution is nevertheless interesting as a benchmark to understand more realistic scenarios, which we shall investigate in the next subsections.

Figure~\ref{fig:successive_shocks} shows the evolution of the CR spectrum for $M_0=20$, $T_0=10^6$~K ($u_0 = 3320$~km/s), $n_0 = 0.01$~cm$^{-3}$, $\xi=3$, $p_{\rm max}=1$~PeV. There are no preexisting particles before the first shock, and particles are injected from the thermal pool at each shock. The bottom right panel displays the effective flow velocity felt by particles of momentum $p$. In Bohm's diffusion regime, the diffusion length is proportional to $p$ such that $x_p$, the distance probed by particles of momentum $p$ ahead of the shock, can be identified with the physical distance up to a rescaling ($x_{p_{\rm max}}$ being interpreted as the position of a free escape boundary). The curves in the bottom right panel can therefore be readily identified with the velocity profiles of the scattering centres. One sees how the energetic particles slow down the flow upstream of the shock, leading to the formation of a precursor. The effective compression ratio of the subshock decreases rapidly after the first few shocks and stabilises around 2.5, which results in a steepening at low energies: the spectral index can be as high as 5. On the other hand, the total compression ratio increases after each shock before it stabilises around 4.6, which leads to a hardening of the high energy bands. Asymptotically, the spectrum is harder than $p^{-4}$ beyond 100~GeV and the spectral index reaches 3.6 at 1~PeV. This demonstrates a striking discrepancy compared with the linear evolution displayed for comparison on the top right panel of Figure~\ref{fig:successive_shocks}. As shown by e.g. \cite{melrose1993}, the linear treatment indeed leads to a spectral hardening until the asymptotic $p^{-3}$ distribution is reached.

\begin{figure*}
\centering
  \includegraphics[width=\linewidth]{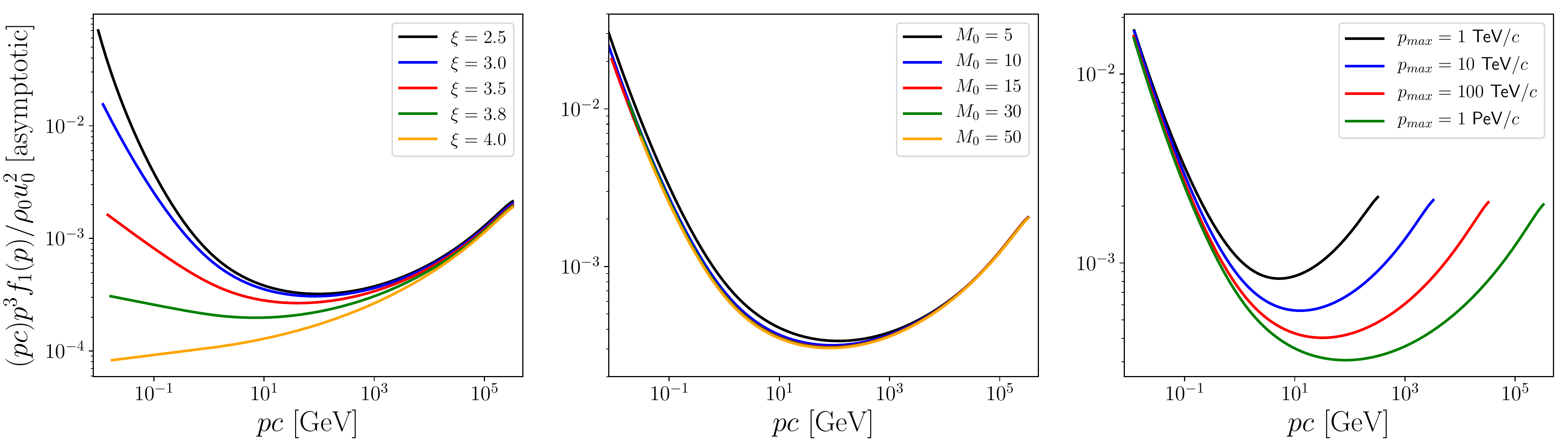}
\caption{Asymptotic solution after multiple shock reaccelerations for various sets of parameters. Left: $M_0=20$, $p_{\rm max} = 10^6$ GeV, varying $\xi$. Middle: $\xi=3$, $p_{\rm max} = 10^6$ GeV, varying $M_0$. Right: $M_0=20$, $\xi=3$, varying $p_{\rm max}$.}
\label{fig:asymptotic_spectra}
\end{figure*}
Figure \ref{fig:asymptotic_spectra} shows the asymptotic solution (typically reached after about 10 reaccelerations) for various sets of parameters. It is striking that the high energy part of the spectrum above 100~GeV displays a somewhat universal shape, that is, a slight concavity, with a spectral index decreasing from about 3.9 to 3.5. As far as the injection parameter is concerned, it only affects the lower part of the spectrum, which is steeper for high injection efficiencies (small $\xi$). Furthermore, the asymptotic solution is nearly independent of the Mach number and, up to a rescaling, of the maximum momentum as well.

\begin{figure}
\centering
  \includegraphics[width=\linewidth]{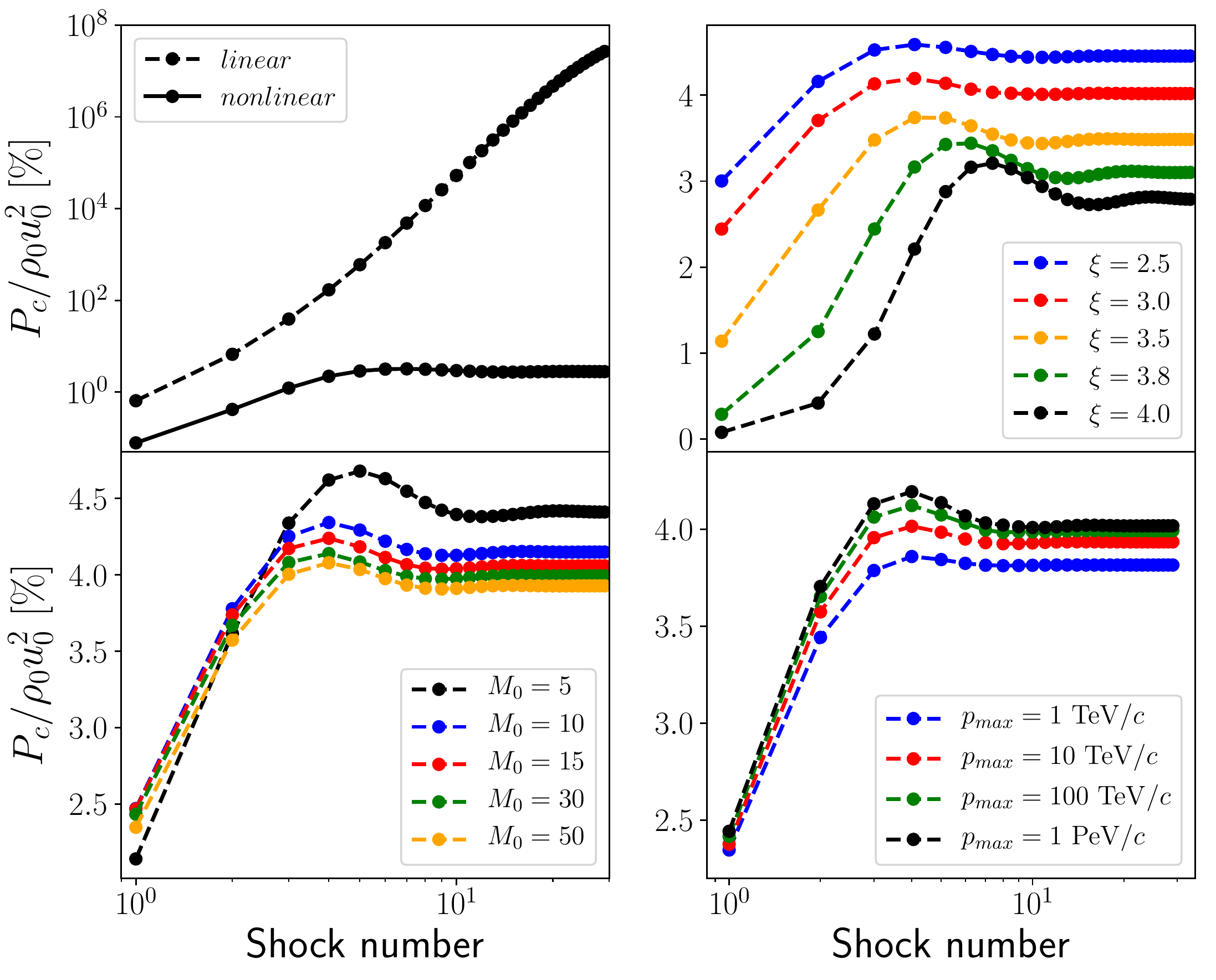}
\caption{Evolution of the CR pressure after reacceleration by successive shocks. Top left panel: comparison between linear and nonlinear computations, for $\mathcal{M}_0=20$, $\xi=4$, $p_{\rm max}=1$~PeV. Top right panel: $\mathcal{M}_0=20$, $p_{\rm max}=1$~PeV, varying $\xi$. Bottom left panel: $\xi=3$, $p_{\rm max}=1$~PeV, varying $M_0$. Bottom right panel: $\mathcal{M}_0=20$, $\xi=3$, varying $p_{\rm max}$.}
\label{fig:evolution_pressure}
\end{figure}
The evolution of the pressure of the postshock CRs (in between the passage of two shocks, after adiabatic decompression) as function of the number of shocks which have already swept-up the medium is plotted in Figure~\ref{fig:evolution_pressure} for various sets of parameters. The comparison between the linear and nonlinear computation displayed on the top left panel demonstrates the need for a nonlinear treatment of the problem in order to avoid energy violation. Noteworthy, with the adopted parameters, the CR pressure calculated in the linear case exceeds 100\% after four shocks, even though we chose a very small injection efficiency ($\xi=4$). The three other panels show that the evolution of the pressure does not depend much on the parameters and can be divided into three phases. At the first three shocks, the pressure increases, until it overshoots its asymptotic value. Then it decreases and stabilises after about 10 shocks. The reason behind the decrease of the downstream pressure is that, as discussed above, the acceleration efficiency decreases when the seed pressure is too large. Eventually a balance is set between the downstream advection and the upstream escape. The asymptotic fluxes are such that about 50\% of the shock kinetic flux goes into CRs, with about 10\% advected downstream and 40\% escaping upstream. This means that the reacceleration process saturates when the system approaches an equipartition of kinetic and thermal energy. This is the very reason behind the nearly universal character of the asymptotic solution. Only the injection of fresh particles modulates the low energy bands of the spectrum and the value of the CR pressure at saturation, which slightly increases as the injection efficiency increases (top right panel of Figure~\ref{fig:evolution_pressure}). This variation is nevertheless very moderate. Only for very low injection fractions (e.g. $10^{-12}$), the pressure of postshock CRs is found to saturate below 1\%. This is because in this case the asymptotic test-particle solution $f(p) \propto p^{-3}$ is reached before nonlinear effects regulate the energy balance. In particular, the energy of the particles always remains negligible compared to that of the shock and equipartition cannot be reached. However such small efficiencies are not realistic, and therefore we conclude that nonlinear effects are unavoidable when dealing with multiple shocks.
In this case, for standard values of the injection parameter ($\xi \sim 2-4$) the pressure of the CRs remaining in the (decompressed) medium is always about $3-5\%$ of the shock ram pressure.


\subsection{Heating}
\begin{figure}
\centering
  \begin{minipage}{0.78\linewidth}
   \centering
   \includegraphics[width=\linewidth]{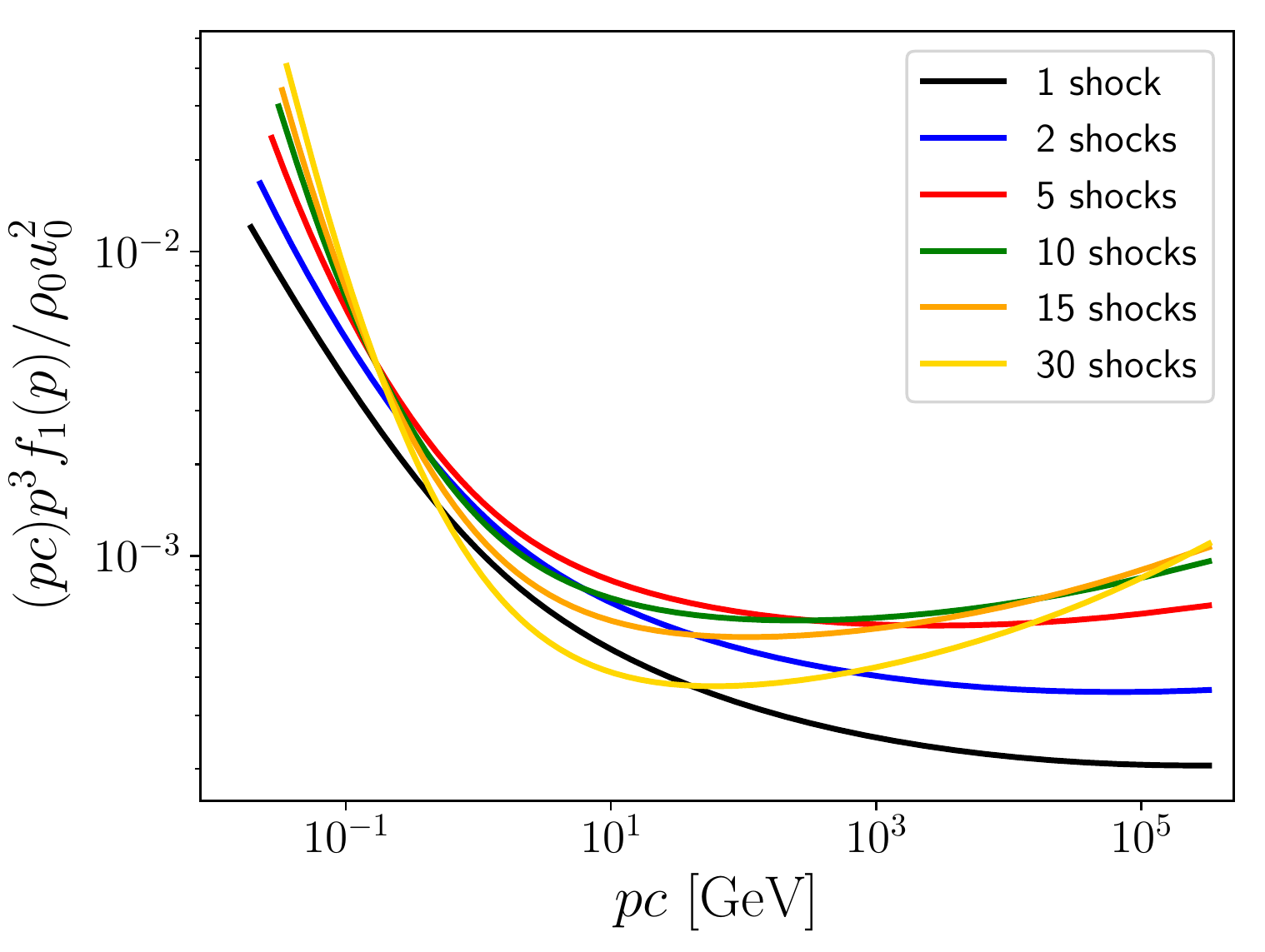} 
  \end{minipage}
\hfill
\centering
  \begin{minipage}{0.82\linewidth}
   \centering ~~~~~~~~
  \includegraphics[width=\linewidth]{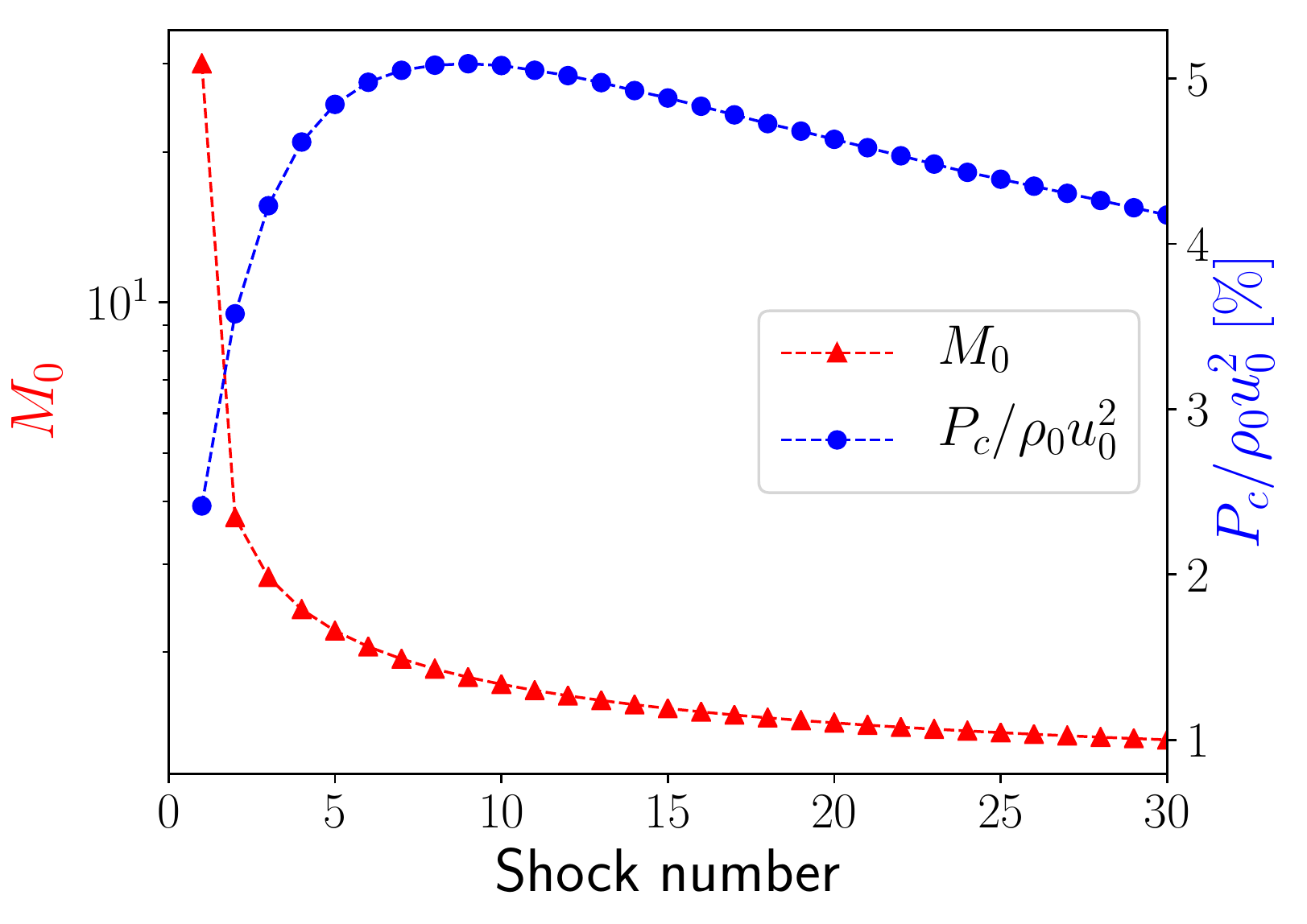}
  \end{minipage}
  \caption{Evolution of the downstream CR spectrum (top panel), the shock Mach number (bottom panel, in red) and the postshock CR pressure (bottom panel, in blue) when the heating of the medium in between successive shocks is taken into account.}
  \label{fig:multipleshocks_variable_M0}
\end{figure}

The spectra discussed in the previous section have been obtained in the idealistic situation where all shocks are identical. This is not expected in realistic environments. For instance, if all shocks span the same volume, the medium is expected to be heated. The postshock temperature including adiabatic decompression reads, as function of the upstream temperature $T_0$ \citep{amato2006}:
\begin{equation}
    T_2 = T_0 R_{\rm sub}^{1-\gamma} (1+\Lambda_B) \frac{\gamma+1+(1-\gamma)/ R_{\rm sub}}{\gamma+1+(1-\gamma) R_{\rm sub}}
    \, ,
\end{equation}
where $\Lambda_B$ has been defined in Eq.~\ref{RHlamb}. In Figure~\ref{fig:multipleshocks_variable_M0}, we plot the evolution of the shock Mach number and postshock CR pressure, accounting for the heating of the medium. The initial temperature is set to $10^6$~K, the shock velocity is fixed to $u_0=5000$~km/s (i.e. the initial Mach number is set to 30) and the injection parameter is set to $\xi=3$.

After a few shocks, the Mach number decreases rapidly and the reacceleration becomes inefficient. The spectra are steeper than in the situation where all shocks are identical. Although an asymptotic solution is also expected to be reached in this case, it takes much longer time for the system to converge.

\subsection{Towards cosmic ray production in superbubbles}
A promising environment where multiple shock acceleration is expected to take place is the interior of a superbubble (SB). Indeed, confined CRs may be successively reaccelerated by SN shocks inside SBs \citep{ferrand2010}. Interestingly, the adiabatic expansion of the SB compensates the heating of the medium due to the successive shocks, such that the interior temperature stays nearly constant in time \citep{parizot2004}. On the other hand, the density is constantly decreasing with time: $n(t) \propto t^{-22/35}$. We computed again the reacceleration of CRs by successive shocks taking into account this density drop with a constant temperature.

\begin{figure}
\centering
  \begin{minipage}{0.8\linewidth}
   \centering
   \includegraphics[width=\linewidth]{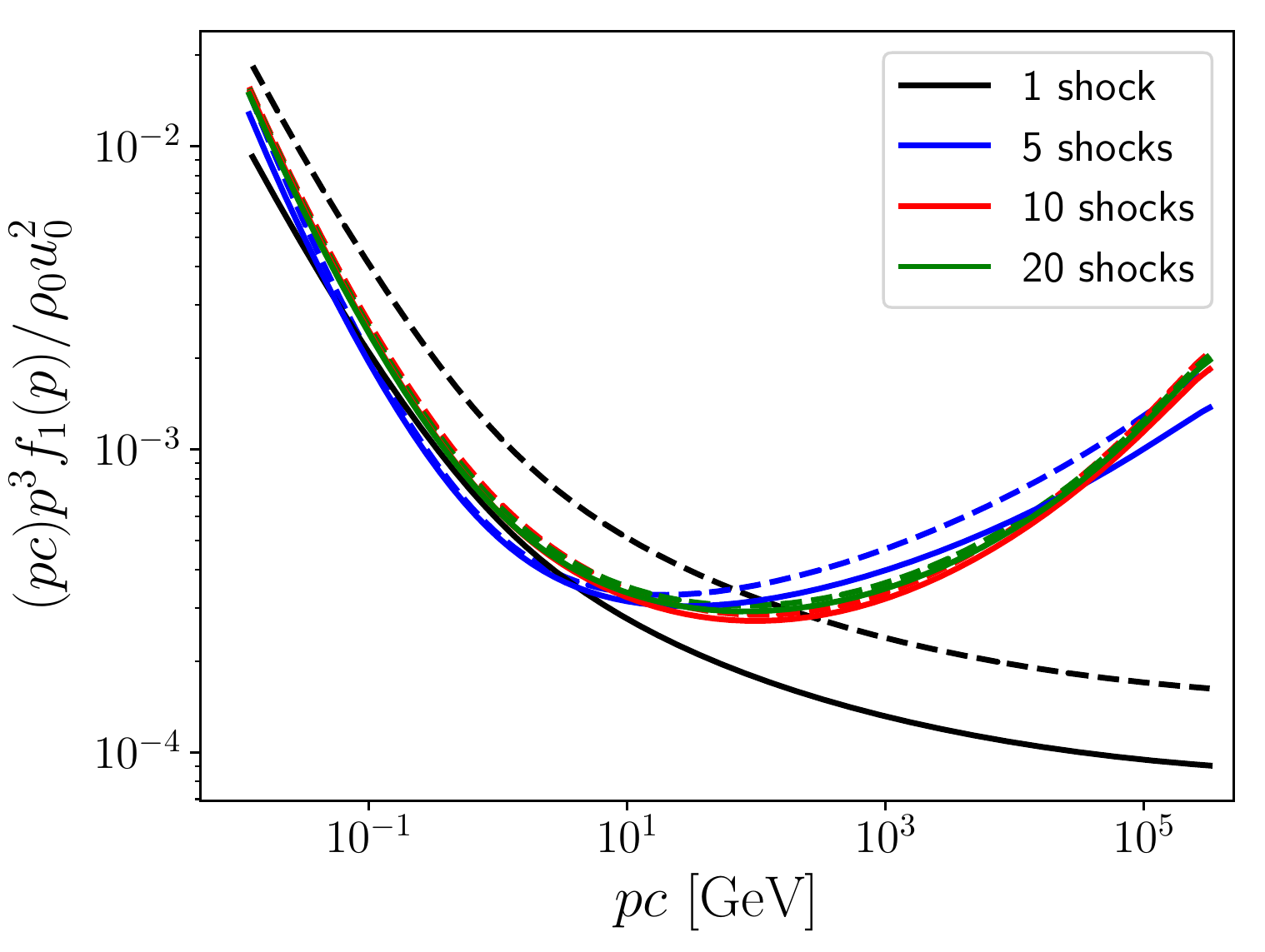} 
  \end{minipage}
\hfill
\centering
  \begin{minipage}{0.8\linewidth}
   \centering
  \includegraphics[width=\linewidth]{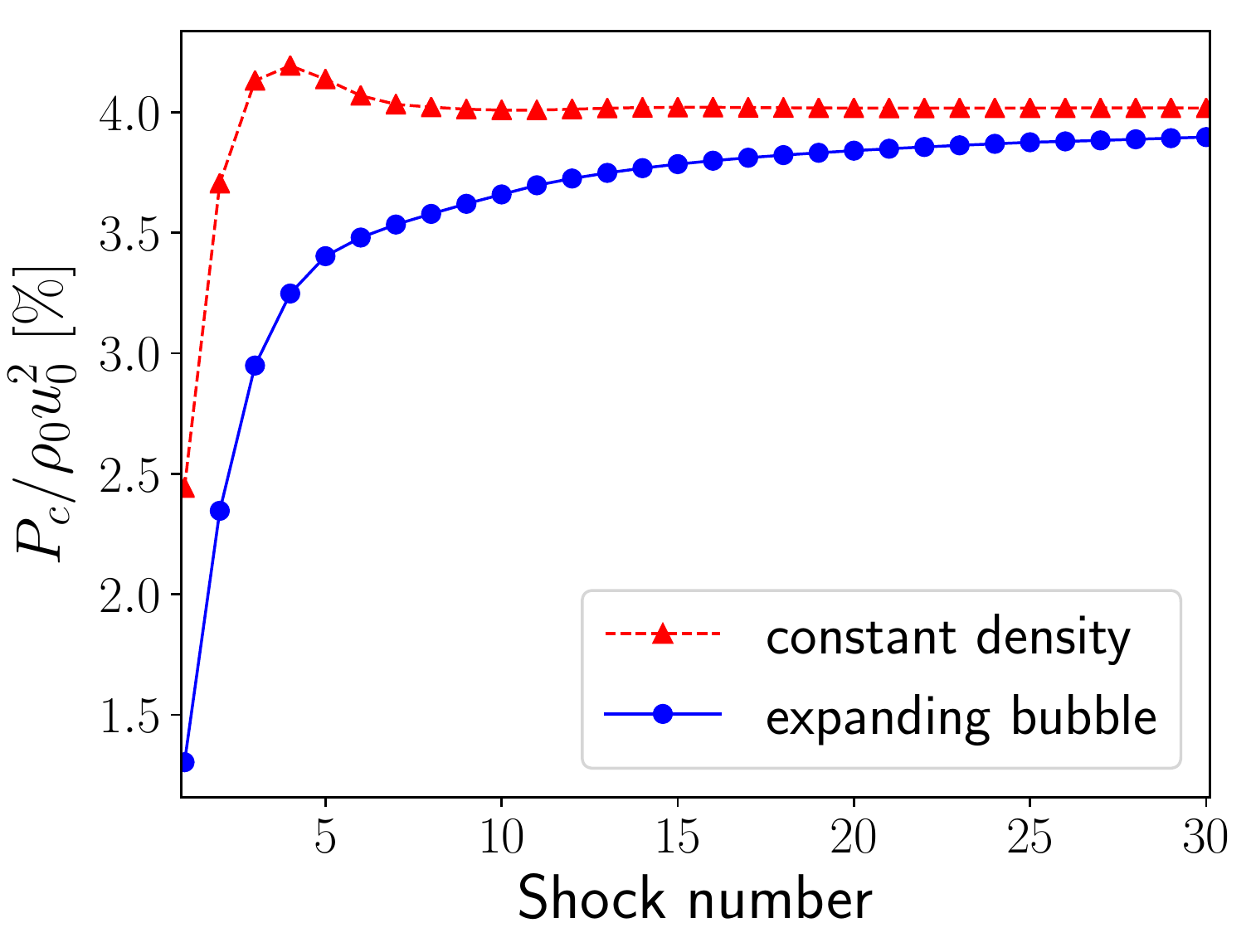}
  \end{minipage}
  \caption{Top panel: Evolution of the CR spectrum during multiple shock acceleration in an expanding medium (solid lines) compared to the spectra computed in the case where all shocks are identical (dashed lines). Bottom panel: Evolution of the CR pressure in an expanding medium. The parameters are $n_0 = 0.01$~cm$^{-3}$, $M_0=20$, $\xi=3$.}
  \label{fig:multipleshocks_superbubbles}
\end{figure}

Figure~\ref{fig:multipleshocks_superbubbles} shows the evolution of the CR spectrum in an environment where the density decreases as $n_i \propto i^{-22/35}$, where $i$ is the number of shocks which have swept-up the medium (we assume that the shocks accelerate the particles at regular intervals), and $n_i$ is the density far upstream of the ith shock. What is shown is the spectrum just before the $(i+1)$th reacceleration (including the decompression of the medium), compared with the postshock spectrum in the case where the ambient density is constant. The bottom panel of Figure~\ref{fig:multipleshocks_superbubbles} displays the evolution of the postshock pressure normalised to the ram pressure of the first shock. Because CRs suffer enhanced adiabatic losses in between shocks, the pressure does not increase as rapidly as in the case of constant density and do not overshoots the asymptotic value. Nevertheless, the asymptotic pressure is identical to that computed in the ideal case of constant density, which is around 4\% of the ram pressure of the first shock. Neither are the particle spectra displaying substantial modifications.

Reducing a SB to an expanding medium is once again a minimalistic approach. In particular, we neglected so far an important aspect of the problem, which is the escape of the particles in between the passage of two shocks. Indeed, in deriving the previous results, it was implicitly assumed that all particles, even that of the highest energies, would stay confined. This may not be the case in realistic environments. For instance, the average time interval between two SN explosions in a typical SB of lifetime 40~Myr is about $\Delta t = 40/N_*~{\rm Myr} \approx 0.1-1$~Myr, where $N_*$ is the number of massive stars in the cluster. On the other hand, the escape time of GeV particles away from the region of acceleration is about 0.01 - 10~Myr, depending on the level of turbulence and the size of the bubble \citep{ICRCtalk,paper3}. In order to probe the modulation induced by the escape of the particles in between the reacceleration events, we assume that the escape time scales as $\tau_{\rm esc}(p) = \tau (p c/1~{\rm GeV})^{-1/3}$, which is expected for relativistic particles in a Kolmogorov turbulence \citep{ferrand2010}. Neglecting all processes but the escape, the transport equation averaged over the SB volume simply reads, in between SN explosions: $\d_t f = -f/\tau_{\rm esc}$, which provides:
\begin{equation}
	f(t) = f(t_i) e^{- \Delta t/\tau (p c/1~{\rm GeV})^{1/3}} \, ,
\end{equation}
where $\Delta t$ is the average time interval between two SN explosions and $f(t_i)$ is the decompressed postshock distribution after the passage of the $i$th shock.

\begin{figure}
\centering
  \begin{minipage}{0.78\linewidth}
   \centering
   \includegraphics[width=\linewidth]{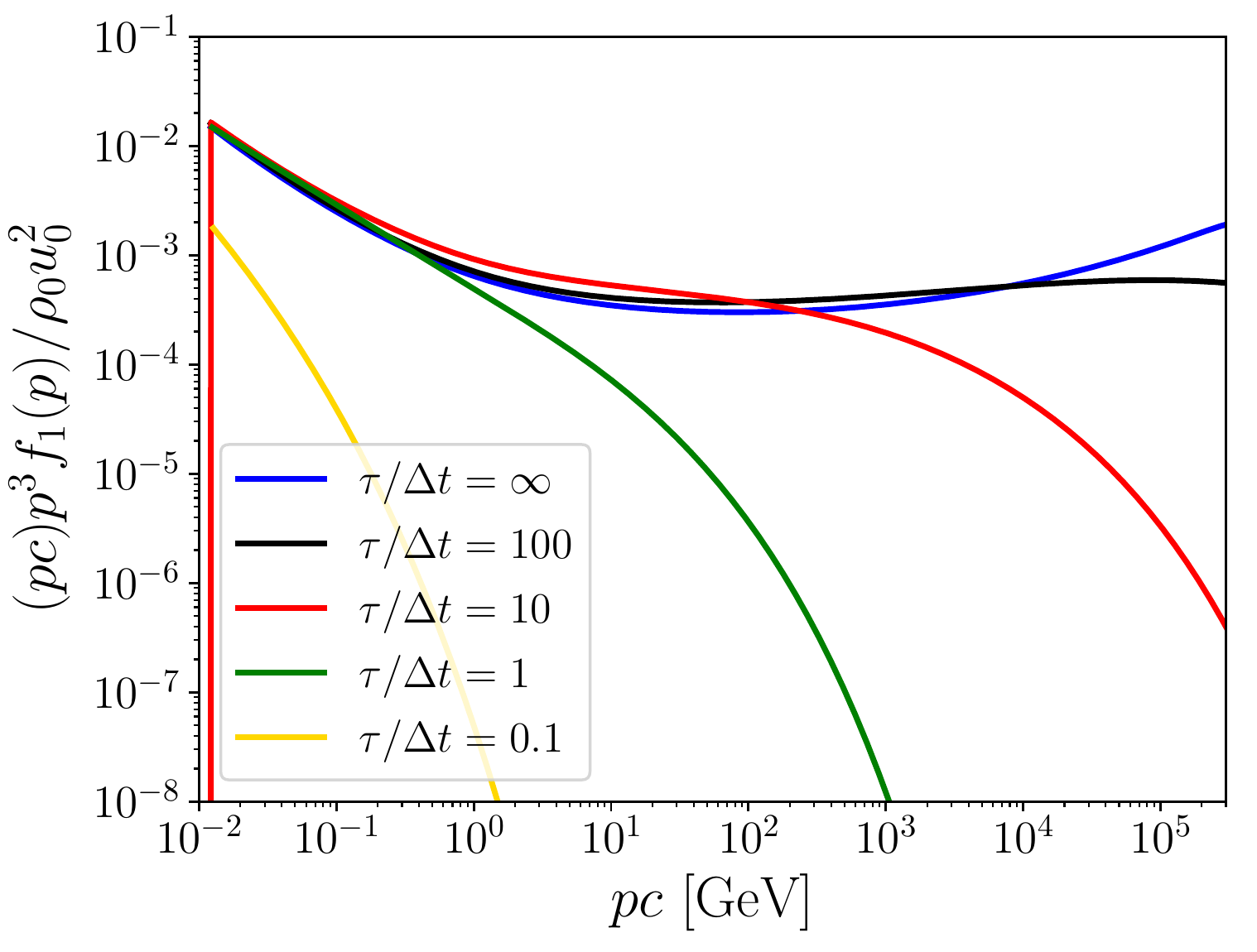} 
  \end{minipage}
\hfill
\centering
  \begin{minipage}{0.78\linewidth}
   \centering
  \includegraphics[width=\linewidth]{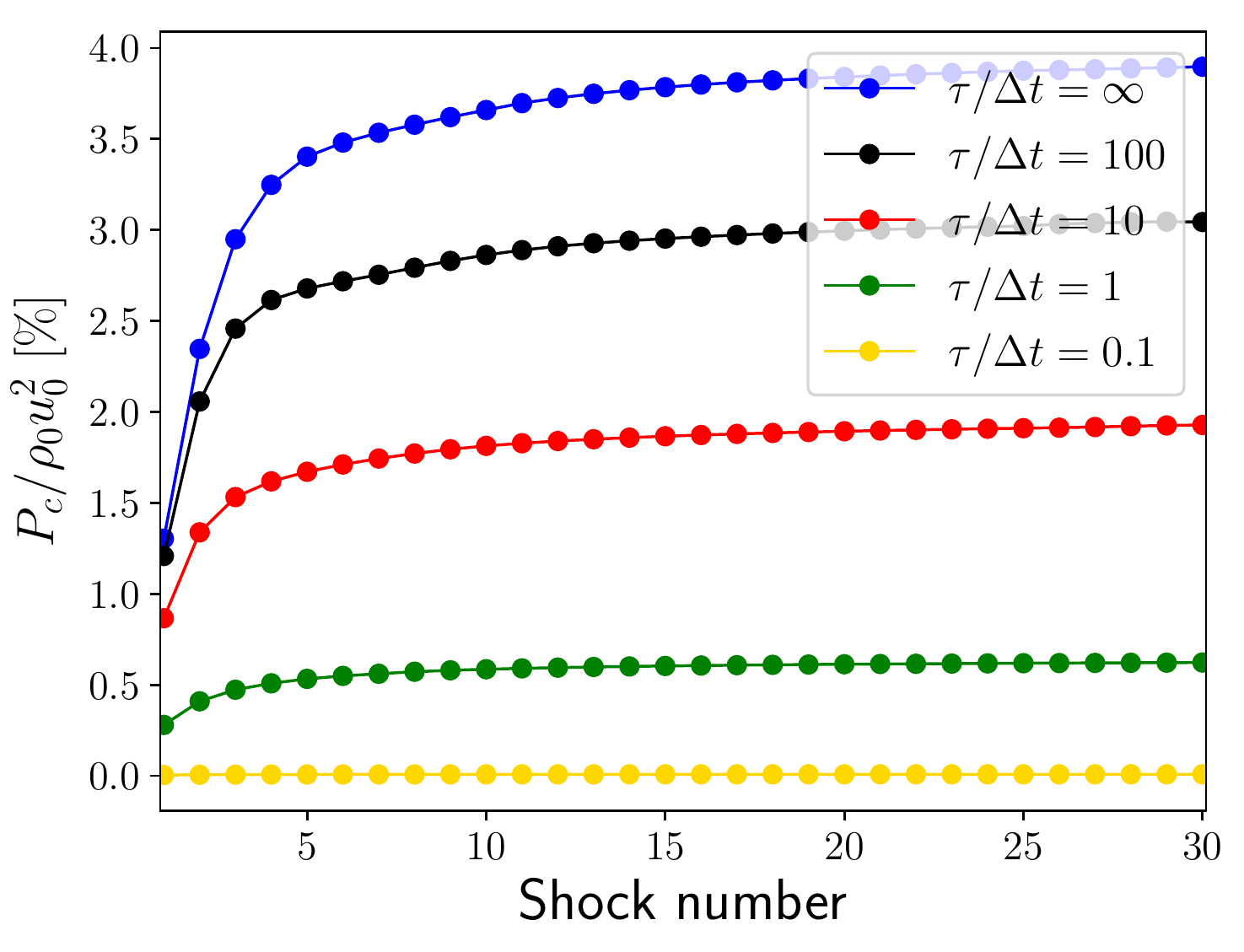}
  \end{minipage}
  \caption{Asymptotic CR spectrum (top) and evolution of the pressure (bottom) during multiple shock accelerations in an expanding medium where the particles can escape in between the passage of two shocks. The parameters are $n_0 = 0.01$~cm$^{-3}$, $M_0=20$, $\xi=3$.
  Each curve corresponds to a different ratio of the particle escape time to the time interval between two SN explosions.}
  \label{fig:multipleshocks_superbubbles_escape}
\end{figure}

Figure~\ref{fig:multipleshocks_superbubbles_escape} shows the resulting CR spectra and pressure evolution. The ``benchmark'' asymptotic solution described in Section~\ref{sec:multipleshocks_identical} is retrieved up to the momentum such that $\tau_{\rm esc} (p) < \Delta t$. Beyond this momentum, the particles escape and are not reaccelerated. With a small escape parameter $\tau$, the CR production may be intermittent at all energy bands. As shown by the yellow curves in Figure~\ref{fig:multipleshocks_superbubbles_escape}, in this case there are nearly no particles remaining in between the shocks and the CR pressure right before a SN explosion is close to zero. The particles are not reaccelerated and the SB is just a collection of isolated SNe.

 On the other hand, the concave asymptotic solution can be retrieved provided the escape parameter is sufficiently large (black curve). This may occur if e.g. the interval between two SNe is $\Delta t \sim 100$~kyr and the escape time is $\tau \sim 10$~Myr. These are not unrealistic values for clusters hosting hundreds of massive stars in possibly very turbulent environments such as the galactic centre. We will soon publish a detailed model of particle acceleration in SBs \citep{ICRCtalk,paper3}, building on the work detailed in the present paper.


\section{Conclusions}\label{secconclusion}
We tackled the problem of the nonlinear reacceleration of particles by a succession of strong shocks, using a semi-analytical computation accounting for the streaming instability as well as the Alfvénic drift effect. We have shown that the linear framework provides a very inaccurate estimate of the solution. The presence of seeds can indeed strongly modify the shock structure and balance the shock pressure such that particle reacceleration becomes less efficient at either low or high energies, depending on the seed spectrum. This can lead to spectra either steep (for steep distributions of seeds), hard (for flat distributions of seeds), or displaying a sharp transition from a steep to a hard component (for hard distributions of seeds).

We then considered the acceleration of particles by multiple identical shocks, in the case where no particles exist before the first shock. The spectrum converges towards an asymptotic solution, typically reached after 20 shocks. Remarkably, the asymptotic spectrum is nearly universal above 10 GeV. In particular, it does not depend on the injection efficiency and shock Mach number. The injection efficiency shapes the low energy bands, which are steeper for higher efficiencies. The asymptotic solution is eventually characterised by a spectral index which increases from around 5 at the injection momentum to about 3.5 at the maximum energy. This is again very different from the linear solution, $f(p) \propto p^{-3}$.

The postshock CR pressure increases after a few shocks, then quickly stabilises around an asymptotic value. The latter is about 4~--~5\% of the ram pressure of one shock. Interestingly, this value weakly depends on the injection efficiency. Even for very small efficiencies, a reacceleration by a few shocks is sufficient for the CR pressure to reach a few percent of the shock energy.

We eventually generalised the analysis to the case of non-identical successive shocks. In a constant volume the medium is heated between each shocks, which leads to a rapid decrease of the shock Mach numbers: the reacceleration of the particles becomes inefficient. On the other hand, assuming an environment undergoing an adiabatic expansion, such as a galactic SB, we found results similar to the case of identical shocks. The particle spectrum converges towards the same asymptotic distribution, with a concave shape and universal spectral index around 3.5 at the maximum energy. Eventually, we allowed the particles to escape the accelerator in between the passage of two shocks, demonstrating that the asymptotic solution would form only in the case where particles are very efficiently confined. This computation will be a building block of a self-consistent model of particle acceleration in SBs \citep{ICRCtalk,paper3}.



\section*{Acknowledgements}
TV acknowledges Alexandre Marcowith and Etienne Parizot for helpful discussions and suggestions. SG and VT acknowledge support from Agence Nationale de la Recherche (grant ANR-17-CE31-0014).




\bibliographystyle{mnras}
\bibliography{biblio} 




\bsp	
\label{lastpage}
\end{document}